\documentclass[12pt, onecolumn]{IEEEtran}
\IEEEoverridecommandlockouts
\usepackage[utf8]{inputenc} 
\usepackage[T1]{fontenc}
\usepackage{url}
\usepackage{ifthen}
\usepackage{cite}
\usepackage[cmex10]{amsmath} 
\usepackage{amssymb}
\usepackage{graphicx}
\usepackage{epstopdf}
\usepackage{epsfig}
\usepackage{multirow}
\usepackage{amsthm}
\usepackage{comment}
\usepackage{caption}
\usepackage{enumitem}
\usepackage{amsmath}
\usepackage{color}
\usepackage{bbm}
\usepackage{algorithm}
\usepackage{algpseudocode}
\usepackage{csquotes}
\usepackage{subfiles}
\usepackage{caption, multirow, makecell}
\usepackage{array}
\usepackage{caption}
\usepackage{afterpage}
\usepackage{dblfloatfix}
\usepackage{hyperref}
\usepackage{booktabs}
\usepackage{tikz}
\usepackage{longtable}
\usepackage{blkarray}
\usepackage{adjustbox}
\newtheorem{thm}{Theorem}
\newtheorem{lem}{Lemma}

\newtheorem{cor}{Corollary}
\newtheorem{example}{Example}

\newtheorem{defn}{Definition}

\def\BibTeX{{\rm B\kern-.05em{\sc i\kern-.025em b}\kern-.08em
		T\kern-.1667em\lower.7ex\hbox{E}\kern-.125emX}}

\begin{document}
	
	\title{Simplified Plotkin-type Bounds for Function-Correcting Codes with $b$-Symbol and Sum-Rank Distances}
	\author{\IEEEauthorblockN{ Sachin Sampath H\IEEEauthorrefmark{1}, N.T. Rashid Ummer\IEEEauthorrefmark{2} and B. Sundar Rajan\IEEEauthorrefmark{1}} \\
		\IEEEauthorblockA{\IEEEauthorrefmark{1}Department of Electrical Communication Engineering, Indian Institute of Science, Bengaluru, India,
		\ \{sachinsampat, bsrajan\}@iisc.ac.in} \\
		\IEEEauthorblockA{\IEEEauthorrefmark{2}Department of Electronics and Communication Engineering, 
			Government Engineering College Wayanad, Kerala, India,
			\ rashidummer@gecwyd.ac.in}}

	\maketitle
	
	\begin{abstract}
		Function-correcting codes (FCCs) are a coding framework that enables reliable communication with minimal redundancy while ensuring the recovery of specific functions of transmitted messages. FCCs have been studied recently under the $b$-symbol distance and the sum-rank distance. Existing general Plotkin-type bounds on the optimal redundancy of FCCs under these distances require computing the pairwise distances between vectors that map to different function values. As a result, these bounds are often computationally difficult to determine. In this paper, we derive simplified Plotkin-type bounds on the optimal redundancy of FCCs under the $b$-symbol and sum-rank distances. These bounds depend only on the number of message vectors corresponding to each function value and the sum of pairwise distances among vectors mapped to the same function value. The proposed bounds apply to arbitrary functions and are considerably easier to evaluate than existing Plotkin-type bounds. The known simplified bounds for linear functions under the $b$-symbol and sum-rank distances are subsumed as special cases of the proposed bounds. Furthermore, we obtain a tighter bound for linear functions under the sum-rank distance than the existing bound.

	\end{abstract}

	
	\section{Introduction}
	Function-correcting codes (FCCs), introduced by Lenz \textit{et al.} in \cite{LBWY}, provide a coding framework for reliably communicating functions of transmitted messages in the presence of channel errors. Unlike conventional error-correcting codes (ECCs), whose objective is the recovery of the entire message, FCCs are designed for applications in which only the value of a prescribed function is required at the receiver. Such scenarios arise in applications such as archival storage, where only certain attributes of the message require additional protection. Compared to conventional ECCs, FCCs can achieve substantially lower redundancy because they are designed to recover only the function value rather than the entire message \cite{LBWY}. The \textit{optimal redundancy} of FCCs is therefore an important parameter of interest. In \cite{LBWY}, Lenz \textit{et al.} established a connection between FCCs and \textit{irregular-distance codes} to derive general bounds on the optimal redundancy in the Hamming distance setting.
	
	FCCs have also been studied under various channel models, such as symbol-pair read channels \cite{XLC}, $b$-symbol read channels \cite{AAY, SR}, channels matched to homogeneous distance \cite{LL}, Lee distance channels \cite{GA, HRS}, insertion-deletion channels \cite{AA}, channels matched to sum-rank distance \cite{SS} and parallel channels \cite{LLM}. A Plotkin-type lower bound on the optimal redundancy of FCCs under the Hamming distance, applicable to arbitrary functions, was proposed in \cite{LBWY}. Another lower bound on the optimal redundancy was obtained in \cite{PR} using function dependent graph. Plotkin-type bounds for the corresponding distance frameworks were also obtained in \cite{XLC, AAY, LL, HRS, AA, SS, LLM}. Each of these Plotkin-type bounds is determined by the distance requirement matrix (DRM) of the target function. Since the entries of the DRM are specified by the distance requirements between message vectors yielding distinct function values, its construction necessitates pairwise distance computations over a large collection of vector pairs.  Obtaining the lower bound using function dependent graph requires the computation of the independence number of the graph, which is also computationally intensive. Simplified lower bounds on the optimal redundancy of FCCs for specific functions, such as the Hamming weight function \cite{LBWY, ZXZG} and the Lee weight function \cite{GA,HRS}, were obtained only for certain parameter regimes. For linear functions, simplified bounds were obtained in \cite{PR,GA,SR,SS} under the Hamming, Lee, $b$-symbol and sum-rank distances, respectively.  
	
	The study of FCCs has been extended in various other directions. The concept of FCCs was extended in \cite{CSFH2} to \textit{function-correcting partition codes}, where the code construction is based on a partition of the message space rather than on a particular target function. In \cite{CSFH3}, another extension known as \textit{FCCs with data protection} was introduced, with the goal of protecting both the original message and the corresponding function value. FCCs with data protection are referred to as \textit{strict FCCs}. Existence and construction of strict FCCs were discussed in \cite{CSFH5}, whereas non-existence of strict FCCs was discussed in \cite{CSFH4}. FCCs with data protection were further studied for a particular class of functions, namely the Hamming code membership function, in \cite{SAARB, AAS}. The concept of \textit{generalized function-correcting partition codes} (GFCPCs) was introduced in \cite{CL} to provide simultaneous protection for multiple partitions of the message space against different error levels.
	
	In this paper, we study FCCs under the $b$-symbol and sum-rank distance frameworks. The $b$-symbol distance was introduced to model communication and storage systems in which the receiver reads overlapping sequences of b consecutive symbols rather than individual symbols. The symbol-pair read channel was first studied by Cassuto and Blaum in \cite{CB}, and was later generalized to the $b$-symbol setting by Yaakobi \textit{et. al.} in \cite{YBS}. The sum-rank metric, on the other hand, generalizes both the Hamming and rank metrics and has found important applications in multishot network coding, distributed storage, and other communication systems. Rank-metric codes were introduced in \cite{KK} and \cite{SKK}. The fundamental properties of sum-rank metric codes and their constructions were discussed in \cite{BG} and \cite{MSK}. FCCs for symbol-pair read channels were studied in \cite{XLC} and later generalized to $b$-symbol read channels in \cite{AAY}. Recently, FCCs for the sum-rank metric were studied in \cite{SS}. The Plotkin-type lower bounds on the optimal redundancy of FCCs developed in \cite{AAY} and \cite{SS} for the respective metrics apply to arbitrary functions. However, these bounds are computationally expensive to evaluate, as they depend on the entries of the DRM associated with the function under consideration. For linear functions, a simplified Plotkin-type bound under the $b$-symbol distance was proposed in \cite{SR}, where the bound depends only on the sum of the Hamming weights of the vectors in the kernel of the function. Similarly, \cite{SS} derived a simplified Plotkin-type bound for linear functions under the sum-rank distance, expressed in terms of the sum-rank weights of the vectors in the kernel of the corresponding function. A simplified lower bound on the optimal redundancy of FCCs for the $b$-symbol weight function was obtained in \cite{AAY}. Therefore, it is of significant interest to derive simplified and computationally tractable bounds on the optimal redundancy of FCCs that are applicable to arbitrary functions under these distance metrics. 
	
	The preimages of a function induce a natural partition of the message space, with each part corresponding to a distinct function value. We exploit this partition to derive simplified Plotkin-type lower bounds on the optimal redundancy of FCCs. The bounds depend only on the cardinalities of the preimage sets and the sums of pairwise distances within each set, thereby avoiding the computation of distances across different sets. We establish Plotkin-type bounds on the optimal redundancy of FCCs for both the $b$-symbol and sum-rank distances, by exploiting the partition of the message space determined by the function. The main contributions of this paper are summarized as follows:
	\begin{itemize}
		\item For arbitrary functions, we derive general Plotkin-type bounds on the optimal redundancy of FCCs under the $b$-symbol and sum-rank distance frameworks. The bounds require only the cardinalities of the function preimages and the sum of distances between pairs of vectors within each preimage, making them substantially simpler to evaluate than existing Plotkin-type bounds based on the entries of the DRM.
		\item We show that the previously known simplified Plotkin-type bounds for surjective linear functions under the $b$-symbol and sum-rank distances can be recovered as special cases of our general bounds. Furthermore, we obtain a tighter bound for linear functions under the sum-rank distance than the existing bound. 
	\end{itemize}

	\subsection{Organization}
	The remainder of the paper is organized as follows. Section \ref{sec_prelim} presents the preliminaries, including the definitions of the $b$-symbol and sum-rank distances, followed by the basic concepts and definitions of FCCs under these distance frameworks. In Section \ref{sec_Plotkin_b}, we derive a general Plotkin-type bound on the optimal redundancy of FCCs under the $b$-symbol distance for arbitrary functions. The bound is then specialized to linear functions. Section \ref{sec_Plotkin_SR} develops the corresponding general Plotkin-type bound under the sum-rank distance and considers its specializations to linear functions. In both sections, we show how the proposed general bounds recover the existing simplified bounds for linear functions. Section \ref{sec_concl} concludes the paper.

	\noindent \textit{Notations:}
	Let $\mathbb{N}_0$ denote the set of non-negative integers and $\mathbb{N}_{0}^{M\times M}$ represents the set of all  $M\times M$ matrices with non-negative integer entries. For a matrix $\mathbf{D}\in\mathbb{N}_{0}^{M\times M}$, we denote its $(i,j)-th$  entry by $[\mathbf{D}]_{ij}$. For an integer $M$, we write $ [M]^+ \triangleq max\{M,0\}$ and for any positive integer $N$, $[N]$ denotes the set $\{1,2,...,N\}$. For any set $\mathcal{A}$, $|\mathcal{A}|$ denotes the cardinality of $\mathcal{A}$.  For sets $\mathcal{A}$ and $\mathcal{B}$, $\mathcal{A} \backslash \mathcal{B}$ denotes the elements in $\mathcal{A}$ but not in $\mathcal{B}$. For a prime power $q$, let $\mathbb{F}_q$ be the finite field with $q$ elements, $\mathbb{F}_q^{m \times n}$ denotes the set of all $m \times n$ matrices with entries from the field $\mathbb{F}_q$.
	
	\section{Preliminaries}\label{sec_prelim}
	In this section, we first review the basic definitions and properties of the $b$-symbol and sum-rank distances. We then review the basic framework of FCCs and extend the relevant definitions to these two distances. 
	\subsection{$b$-symbol distance}
	Motivated by limitations in the reading process of high-density data storage systems, Cassuto and Blaum introduced the framework of symbol-pair read channels in \cite{CB}, where the channel output consists of overlapping pairs of symbols. This framework was later extended to $b$-symbol read channels in \cite{YBS}, where the read channel output consists of $b$ consecutive symbols. We first provide the definition of $b$-symbol read vector. 
	\begin{defn}[$b$-symbol read vector \cite{YBS}]
		 The $b$-symbol read vector corresponding to the vector $\mathbf{x}=(x_0,x_1,...,x_{n-1}) \in \mathbb{F}_q^n$ is defined as 
		 $$\pi_b(\mathbf{x}) = [ (x_0, x_1, \ldots, x_{b-1}), (x_1, x_{2}, \ldots, x_{b}), \ldots , (x_{n-1}, x_0, \ldots, x_{b-2}) ].$$
	\end{defn}
	The $b$-symbol distance between two vectors is defined as follows. 
	\begin{defn}[$b$-symbol distance \cite{YBS}]
		The $b$-symbol distance between two vectors $\mathbf{x}, \mathbf{y} \in \mathbb{F}_q^n$ is defined as 
		$$ d_b(\mathbf{x},\mathbf{y}) = d_H(\pi_b(\mathbf{x}), \pi_b(\mathbf{y})),$$ where $d_H(\mathbf{x},\mathbf{y})$ denotes the Hamming distance between the vectors $\mathbf{x}$ and $\mathbf{y}$. 
	\end{defn}
	   The following example illustrates the $b$-symbol distance.
	   \begin{example}
	   	Consider two vectors in $\mathbb{F}_2^6$,
	   	$\mathbf{x}=(0,1,1,0,1,0), \mathbf{y}=(0,1,0,1,1,0),$
	   	and let $b=3$. The corresponding $3$-symbol vectors are
	   	$\pi_3(\mathbf{x})=[ (0,1,1),(1,1,0),(1,0,1), (0,1,0),(1,0,1),(0,1,0) ]$ and $\pi_3(\mathbf{y})=[ (0,1,0),(1,0,1),(0,1,1),(1,1,0),(1,0,1),(0,1,0)]$.
	   	The $3$-symbol distance is the Hamming distance between the corresponding $3$-symbol vectors. Hence, $d_3(\mathbf{x},\mathbf{y})=4.$
	   \end{example}
	    For a code $\mathcal{C}$ over $\mathbb{F}_q$, the $b$-symbol read code of $\mathcal{C}$ is a code over  $\mathbb{F}_q^b$ given by
	    $\pi_b(\mathcal{C}) = \{\pi_b(\boldsymbol{c}) : \boldsymbol{c} \in \mathcal{C}\} $. The minimum $b$-distance of the code, $d_b(\mathcal{C})$, is given by  $d_b(\mathcal{C}) = d_H (\pi_b(\mathcal{C})),$
	    where $d_H (\pi_b(\mathcal{C}))$ is the minimum Hamming distance of the code $\pi_b(\mathcal{C})$. A code $\mathcal{C}$ can correct any $t$ $b$-symbol errors if  $d_b(\mathcal{C}) \geq 2t + 1$ \cite{YBS}. For further details on $b$-symbol distance, readers can refer to \cite{YBS}.

	\subsection{Sum-rank distance}
	The sum-rank distance is a natural generalization of both the Hamming and rank distances. The basic definitions related to sum-rank distance is reviewed in this subsection. 
	\begin{defn}[Sum-rank distance \cite{BG}]
		Let $q$ be a prime power and $t, n_1,...,n_t, m_1,...,m_t \ge 1 $ be integers such that $n_i \le m_i$ for all $i \in [t]$ and $m_1 \ge ... \ge m_t$. Consider the Cartesian product of $t$ matrix spaces given by $\mathbf{\Pi}= \mathbb{F}_q^{m_1 \times n_1}\times...\times \mathbb{F}_q^{m_t \times n_t}$. The sum-rank weight of an element $\mathbf{X}=(\mathbf{X}_1,...,\mathbf{X}_t) \in \mathbf{\Pi}$ is defined as $srk(\mathbf{X}) = \displaystyle \sum_{i=1}^{t} rk(\mathbf{X}_i)$, where $rk(\mathbf{M})$ denotes the rank of $\mathbf{M}$.  The sum-rank distance between two elements $\mathbf{X}=(\mathbf{X}_1,...,\mathbf{X}_t), \mathbf{Y}=(\mathbf{Y}_1,...,\mathbf{Y}_t) \in \mathbf{\Pi}$ is defined as $d_{srk}(\mathbf{X},\mathbf{Y}) = srk(\mathbf{X}-\mathbf{Y})$.
	\end{defn}
	If $m_i =1$ for all $ i \in [t]$, then $\mathbf{\Pi}= \mathbb{F}_q^n$, where $n=n_1+...+n_t$ and the sum-rank distance coincides with the Hamming distance. On the other hand, if $t=1$, then $\mathbf{\Pi}= \mathbb{F}_q^{m \times n}$ and the sum-rank distance coincides with the rank distance on $\mathbf{\Pi}$. The following example illustrates the sum-rank distance.
	\begin{example}
		Consider the case $t=2$ with $m_1=m_2=n_1=n_2=2$. Let
		$
		\mathbf{X}=(\mathbf{X}_1,\mathbf{X}_2),
		\mathbf{Y}=(\mathbf{Y}_1,\mathbf{Y}_2),
		$
		where
		\[
		\mathbf{X}_1=
		\begin{pmatrix}
			1&0\\
			0&1
		\end{pmatrix},
		\quad
		\mathbf{Y}_1=
		\begin{pmatrix}
			1&1\\
			0&1
		\end{pmatrix},
		\qquad
		\mathbf{X}_2=
		\begin{pmatrix}
			1&0\\
			1&0
		\end{pmatrix},
		\quad
		\mathbf{Y}_2=
		\begin{pmatrix}
			0&0\\
			1&0
		\end{pmatrix}.
		\]
		Then
		\[
		\mathbf{X}_1-\mathbf{Y}_1=
		\begin{pmatrix}
			0&-1\\
			0&0
		\end{pmatrix},
		\qquad
		\mathbf{X}_2-\mathbf{Y}_2=
		\begin{pmatrix}
			1&0\\
			0&0
		\end{pmatrix},
		\]
		and hence
		\[
		rk(\mathbf{X}_1-\mathbf{Y}_1)
		=
		rk(\mathbf{X}_2-\mathbf{Y}_2)=1.
		\]
		Therefore, the sum-rank distance between $\mathbf{X}$ and $\mathbf{Y}$ is
		\[
		d_{srk}(\mathbf{X},\mathbf{Y})
		=
		rk(\mathbf{X}_1-\mathbf{Y}_1)
		+
		rk(\mathbf{X}_2-\mathbf{Y}_2)
		=1+1=2.
		\]
	\end{example}
	For a code $\mathcal{C} \subseteq \mathbf{\Pi}$, the minimum sum-rank distance of $\mathcal{C}$, denoted by $d_{srk}(\mathcal{C})$, is defined as	$$d_{srk}(\mathcal{C}) = \displaystyle \min_{\substack{\mathbf{X},\mathbf{Y}\in\mathcal{C}\\
			\mathbf{X}\neq\mathbf{Y}}} d_{srk}(\mathbf{X},\mathbf{Y}).$$
	Consider the special class of codes when $m_i =m, \forall i \in [t]$. In this case, $\mathbf{\Pi} \overset{\sim}{=} \mathbb{F}_{q^m}^N$, where $N=\displaystyle \sum_{i=1}^t n_i$. A code $\mathcal{C}$ in $\mathbb{F}_{q^m}^N$  can correct any $t$ sum-rank errors whenever $d_{srk}(\mathcal{C}) \geq 2t+1$ \cite{NU}. For further details on sum-rank distance, readers can refer to \cite{KK,SKK,BG,MSK}.
	
 In this paper, we consider FCC for sum-rank distance codes over $\mathbf{\Pi} \overset{\sim}{=}(\mathbb{F}_q^{m \times m})^k$, where $(\mathbb{F}_q^{m \times m})^k$ denotes the set of all $k$ length vectors of $m \times m$ matrices. 
 
	Next, we review the basic concepts and definitions of FCCs under the $b$-symbol and sum-rank distances. Although the FCC frameworks under the $b$-symbol and sum-rank distances are analogous, we state their definitions explicitly for completeness and convenience.
	\subsection{Function-correcting codes under the $b$-symbol distance}
	Let $\boldsymbol{x}=(x_0,x_1,\ldots,x_{k-1})\in\mathbb{F}_q^k$ be a
	message vector. Let $f:\mathbb{F}_q^k\rightarrow Im(f)$ be the function of interest, where $Im(f)=\{f(\boldsymbol{x}):\boldsymbol{x}\in\mathbb{F}_q^k\}.$ 
	The cardinality of $Im(f)$, denoted by $E=|\operatorname{Im}(f)|$, is referred to as the expressiveness of the function $f$. The message vector $\boldsymbol{x}$ is encoded using a systematic encoding function
	$\operatorname{Enc}:\mathbb{F}_q^k\longrightarrow\mathbb{F}_q^{k+r},
	\operatorname{Enc}(\boldsymbol{x}) = \bigl(\boldsymbol{x},p(\boldsymbol{x})\bigr),$
	where $p(\boldsymbol{x})\in\mathbb{F}_q^r$ denotes the redundancy
	associated with $\boldsymbol{x}$, and $r$ is the redundancy length.
	The resulting codeword is transmitted through a noisy $b$-symbol read
	channel. We assume that the channel can introduce at most $t$ errors
	in a codeword, where $t>\left\lfloor\frac{b-1}{2}\right\rfloor$.
	Under this setting, a function-correcting $b$-symbol code for $f$ is
	defined as follows. 
	\begin{defn}[Function-Correcting $b$-symbol Codes\cite{AAY}]\label{FCC_defn}
		An encoding function, $ Enc: {\mathbb{F}^k_q} \rightarrow {\mathbb{F}_q^{k+r}},  Enc(\boldsymbol{x}) = (\boldsymbol{x},p(\boldsymbol{x}))$ defines a function correcting $b$-symbol code (FCBC) for the function $f: \mathbb{F}^k_q \rightarrow Im(f)$  if 
		\begin{equation*}  
			d_b(Enc(\boldsymbol{x_1}), Enc(\boldsymbol{x_2})) \geq 2t+1 \hspace{3mm}\forall \quad\boldsymbol{x_1}, \boldsymbol{x_2} \in {\mathbb{F}^k_q} \text{ with  } f(\boldsymbol{x_1}) \neq f(\boldsymbol{x_2}).  
		\end{equation*}
	\end{defn}
	Since the primary objective of an FCBC is to recover the desired function value with the least possible redundancy, the optimal redundancy is defined as follows.
	
	\begin{defn}[Optimal Redundancy of an FCBC \cite{AAY}]
		The smallest positive integer $r$ for which an FCBC exists with an encoding function $ Enc: {\mathbb{F}^k_q} \rightarrow {\mathbb{F}_q^{k+r}}$ for the function $f$ is called the \textbf{optimal redundancy} of $f$.  It is denoted as $r_b^f(k,t)$.
	\end{defn}
	Two types of $b$-symbol distance requirement matrices were defined in \cite{AAY} and are stated below.
	\begin{defn}[$b$-symbol Distance Requirement Matrices \cite{AAY}] 
		Consider $M$ vectors $\boldsymbol{x_1}, . . . , \boldsymbol{x_{M}} \in {\mathbb{F}^k_q}$. Then,  $\boldsymbol{B_f^{(1)}}(t, \boldsymbol{x_1}, . . . , \boldsymbol{x_{M}})$ and
		$\boldsymbol{B_f^{(2)}}(t, \boldsymbol{x}_1, . . . , \boldsymbol{x}_{M})$  are $M \times M$  $b-$symbol distance requirement matrices (BDRMs) with entries
		
		\begin{equation}\label{BDRM1}
			[\boldsymbol{B_f^{(1)}}(t, \boldsymbol{x}_1, . . . , \boldsymbol{x}_{M})]_{ij}=   
			\begin{cases}
				[2t - b + 2 - d_b(\boldsymbol{x}_i, \boldsymbol{x}_j)]^+ , & if \hspace{1mm} f (\boldsymbol{x}_i) \neq f(\boldsymbol{x}_j),\\0, & otherwise.
			\end{cases}
		\end{equation}
		
		and 
		
		\begin{equation}
			[\boldsymbol{B_f^{(2)}}(t, \boldsymbol{x}_1, . . . , \boldsymbol{x}_{M})]_{ij}=   
			\begin{cases}
				[2t + b - d_b(\boldsymbol{x}_i, \boldsymbol{x}_j)]^+ , & if \hspace{1mm} f (\boldsymbol{x}_i) \neq f(\boldsymbol{x}_j),\\0, & otherwise.
			\end{cases}
		\end{equation}
	\end{defn}
	Given a matrix $\mathbf{B} \in \mathbb{N}_0^{M \times M}$, a $b$-symbol distance code is defined as follows.  
	    \begin{defn}[$\boldsymbol{B_b}$-code \cite{AAY}] A set of codewords $\mathcal{P} = \{ \boldsymbol{p_1}, \ldots, \boldsymbol{p_M} \}$ is said to be a $\boldsymbol{B}$-irregular $b-$symbol distance code ($\boldsymbol{B_b}$-code) for some matrix $\boldsymbol{B}$ $\in \mathbb{N}_0^{M \times M}$ if there exists an ordering of the codewords in $\mathcal{P}$ such that $d_b(\boldsymbol{p_i},\boldsymbol{p_j}) \geq$ $\boldsymbol{[B]_{ij}}$ for all $i,j \in \{1, 2, \ldots, M\}.$
	\end{defn}
	The smallest integer $r$  for which there exists a $\boldsymbol{B_b}-$code of length $r$ is denoted by $N_b(\boldsymbol{B})$. As proved in \cite{AAY}, for any function $f: \mathbb{F}^k_q \rightarrow Im(f)$ and $\{\boldsymbol{x_1}, \ldots, \boldsymbol{x_{q^k}}\} = \mathbb{F}^k_q$, we have 
	\begin{equation}
	N_b(\boldsymbol{B}_f^{(1)}(t)) \leq r_b^f(k,t) \leq N_b(\boldsymbol{B}_f^{(2)}(t)).
	\end{equation}
	The following example illustrates the definitions of BDRM and $\mathbf{B}_b$-code.
	\begin{example}
			Let $q=2$, $k=4$, $b=3$, and $t=3$. Consider the four vectors
			\[
			\boldsymbol{x}_1=(0,0,0,0),\quad
			\boldsymbol{x}_2=(0,0,0,1),\quad
			\boldsymbol{x}_3=(0,0,1,1),\quad
			\boldsymbol{x}_4=(0,1,1,1)
			\]
	and define the function $f$ by
				$f(\boldsymbol{x}_1)=0, f(\boldsymbol{x}_2)=1,  f(\boldsymbol{x}_3)=1 \text{ and } f(\boldsymbol{x}_4)=0$.			
				The pairwise $3$-symbol distances among these vectors are
				\[
				\left(d_3(\boldsymbol{x}_i,\boldsymbol{x}_j)\right)_{i,j}
				=
				\begin{bmatrix}
					0&3&4&4\\
					3&0&3&4\\
					4&3&0&3\\
					4&4&3&0
				\end{bmatrix}.
				\]
				Since $b=3$ and $t=3$, we have $2t-b+2=5$ and $2t+b=9$.
				Therefore, the first BDRM has entries
				\[
				[\boldsymbol{B}_f^{(1)}]_{ij}
				=
				\begin{cases}
					[5-d_3(\boldsymbol{x}_i,\boldsymbol{x}_j)]^+,
					& f(\boldsymbol{x}_i)\neq f(\boldsymbol{x}_j),\\
					0,&\text{otherwise}.
				\end{cases}
				\text{ Hence, }
				\boldsymbol{B}_f^{(1)}
				=
				\begin{bmatrix}
					0&2&1&0\\
					2&0&0&1\\
					1&0&0&2\\
					0&1&2&0
				\end{bmatrix}.
				\]
				
				Similarly, the second BDRM is given by
				\[
				[\boldsymbol{B}_f^{(2)}]_{ij}
				=
				\begin{cases}
					[9-d_3(\boldsymbol{x}_i,\boldsymbol{x}_j)]^+,
					& f(\boldsymbol{x}_i)\neq f(\boldsymbol{x}_j),\\
					0,&\text{otherwise},
				\end{cases}
				\text{ which gives }
				\boldsymbol{B}_f^{(2)}
				=
				\begin{bmatrix}
					0&6&5&0\\
					6&0&0&5\\
					5&0&0&6\\
					0&5&6&0
				\end{bmatrix}.
				\]
				Consider the BDRM $\boldsymbol{B}_f^{(2)}$.	A $\boldsymbol{B}_3$-code corresponding to this matrix can be
				constructed by taking codewords of length $6$ as
				$\boldsymbol{p}_1=000000, \boldsymbol{p}_2=111111,\boldsymbol{p}_3=000111,\text{ and } \boldsymbol{p}_4=111000$.
				The corresponding pairwise $3$-symbol distance matrix is
				\[
				\left(d_3(\boldsymbol{p}_i,\boldsymbol{p}_j)\right)_{i,j}
				=
				\begin{bmatrix}
					0&6&5&5\\
					6&0&5&5\\
					5&5&0&6\\
					5&5&6&0
				\end{bmatrix}.
				\]
				Clearly, $\left(d_3(\boldsymbol{p}_i,\boldsymbol{p}_j)\right)_{i,j}
				\geq
				\boldsymbol{B}_f^{(2)}$. Hence, $\mathcal{P}=
				\{\boldsymbol{p}_1,\boldsymbol{p}_2,\boldsymbol{p}_3,\boldsymbol{p}_4\}$ is a $\boldsymbol{B}_3$-code of length $6$ corresponding to $\boldsymbol{B}_f^{(2)}$.
	\end{example}
	The following lemma will be used in the derivation of the general Plotkin-like bound in Section \ref{sec_Plotkin_b}.
	\begin{lem}\cite{AAY}\label{lem:Plotkin}
		Let $\mathcal{P} = \{ \mathbf{p}_1, \mathbf{p}_2, \ldots, \mathbf{p}_M \} \subseteq \mathbb{F}_q^r$ be a code of length $r$ and size $M$ under the b-symbol distance $d_b$. Then, we have
		\begin{equation}\label{eq:Plotkin}
			\displaystyle \sum_{ i,j \in [M] : i <j} d_b(\mathbf{p}_i,\mathbf{p}_j) \le  
			\begin{cases}
				\frac{(q^b - 1)M^2r}{2q^b} , & \text{if }  M \equiv 0 \pmod{q^b}, \\[12pt]
				\frac{(q^b - 1)(M^2 - 1)r}{2q^b}, & \text{if } M \equiv 1 \pmod{q^b}, \\[12pt]
				\frac{(M^2(q^b - 1) - m(q^b - m))r}{2q^b}, & \text{if } M \equiv m \pmod{q^b},\text{ } m>1.
			\end{cases}
		\end{equation} 
	\end{lem}
	
	\subsection{Function-correcting codes under the sum-rank distance}
	Let $\mathbf{U}\in(\mathbb{F}_q^{m\times m})^k$ denote the message, and let $f:(\mathbb{F}_q^{m\times m})^k\longrightarrow \operatorname{Im}(f)$
	be the function to be computed, where $\operatorname{Im}(f)=
	\{f(\mathbf{U}):\mathbf{U}\in(\mathbb{F}_q^{m\times m})^k\}$ and $E=|\operatorname{Im}(f)|.$
	The message is encoded using the systematic encoding function
	$\mathrm{Enc}:(\mathbb{F}_q^{m\times m})^k \longrightarrow
	(\mathbb{F}_q^{m\times m})^{k+r},$ given by $\mathrm{Enc}(\mathbf{U}) = (\mathbf{U},\mathbf{p}(\mathbf{U}))$,
	where $\mathbf{p}(\mathbf{U})\in(\mathbb{F}_q^{m\times m})^r$
	is the redundancy part and $r$ denotes the redundancy. The channel may introduce errors into the transmitted codeword, resulting in an output
	$\mathbf{Y}\in(\mathbb{F}_q^{m\times m})^{k+r}$ satisfying
	$d_{\mathrm{srk}}
	\bigl(\mathrm{Enc}(\mathbf{U}),\mathbf{Y}\bigr)\leq t$.
	The formal definition of a function-correcting code under the
	sum-rank metric, referred to as function-correcting sum-rank codes, is given below.
	\begin{defn}[Function-Correcting Sum-Rank Codes \cite{SS}]
		\label{Function-Correcting Sum-Rank Codes}
		An encoding function \( \mathrm{Enc} : (\mathbb{F}_q^{m \times m})^k \to (\mathbb{F}_q^{m \times m})^{k+r} \),  
		$ \mathrm{Enc}(\mathbf{U}) = (\mathbf{U}, \mathbf{p}(\mathbf{U})) $ defines a function-correcting sum-rank code (FCSRC) for the function $ f : (\mathbb{F}_q^{m \times m})^k \to Im(f) $  if for all $ \mathbf{U}_1, \mathbf{U}_2 \in (\mathbb{F}_q^{m \times m})^k $ with $ f(\mathbf{U}_1) \neq f(\mathbf{U}_2)$, we have $$d_{srk}(\mathrm{Enc}(\mathbf{U}_1), \mathrm{Enc}(\mathbf{U}_2)) \geq 2t + 1.$$ 
	\end{defn}
	The optimal redundancy of an FCSRC associated with a function $f$ is defined as follows.
	\begin{defn}[Optimal Redundancy of an FCSRC \cite{SS}]
		\label{Optimal Redundancy}
		The optimal redundancy \( r_{srk}^f(m, k, t) \) of an FCSRC is defined as the smallest integer \( r \) for which there exists an FCSRC with an encoding function \( \mathrm{Enc} : (\mathbb{F}_q^{m \times m})^k \to (\mathbb{F}_q^{m \times m})^{k+r} \) that enables recovery of \( f(\mathbf{U}) \) under $t$ sum-rank errors.
	\end{defn}
	The sum-rank distance requirement  matrix and the $\mathbf{D}$-sum-rank-distance code associated with a function \(f\) are defined as follows.
	\begin{defn}[Sum-rank Distance Requirement  Matrix \cite{SS}] 
		\label{DRM}
		Let $\mathbf{U}_1, \dots, \mathbf{U}_M \in (\mathbb{F}_q^{m \times m})^k$. The sum-rank distance requirement  matrix (SDRM) 	$\mathbf{D}_{srk}^f(t, \mathbf{U}_1, \dots, \mathbf{U}_M)$ of a function $f$ is defined as the $M \times M$ matrix with entries
		\begin{equation}\label{SDRM}
		[\!\mathbf{D}_{srk}^f\!(t,\! \mathbf{U}_1, \dots, \mathbf{U}_M)\!]_{ij}\!=\!
		\begin{cases}
			\begin{aligned}
				&\![\!2t\! +\!1\!-\!d_{srk}\!(\!\mathbf{U}_i,\! \mathbf{U}_j\!)\!]^+\!, \text{if }\! f(\!\mathbf{U}_i\!)\!\neq\!f(\!\mathbf{U}_j\!), \\
				&0, \quad \text{otherwise}.
			\end{aligned}
		\end{cases}
		\end{equation}
	\end{defn}
	\begin{defn}[$\mathbf{D}_{srk}$-code]
		\label{D-code}
		Let $\mathbf{D} \in \mathbb{N}_0^{M \times M}$ and $\mathcal{P} = \{ \mathbf{P}_1, \mathbf{P}_2, \ldots, \mathbf{P}_M \} \subseteq (\mathbb{F}_q^{m \times m})^k$ be a code of length $r$ and cardinality $M$. Then, $\mathcal{P} = \{ \mathbf{P}_1, \mathbf{P}_2, \ldots, \mathbf{P}_M \}$ is a $\mathbf{D}$-irregular-sum-rank-distance code ($\mathbf{D}_{srk}$-code), if there exists an ordering of the codewords of $\mathcal{P}$ such that $d_{srk}(\mathbf{P}_i, \mathbf{P}_j) \geq [\mathbf{D}]_{ij}$ for all $i, j \in [M]$.
	\end{defn}
	We denote by $N_{srk}(\mathbf{D})$ the smallest length $r$ for which a $\mathbf{D}_{srk}$-code of length $r$ exists. For any function $f : (\mathbb{F}_q^{m \times m})^k \to \mathrm{Im}(f)$, the optimal redundancy satisfies $r_{srk}^f(m,k,t)\;=\;N_{srk}(\mathbf{D}_{srk}^f(t; \mathbf{U}_1, \ldots, \mathbf{U}_{(q^{m \times m})^{k}})),$ where $\{\mathbf{U}_1, \ldots, \mathbf{U}_{(q^{m \times m})^{k}}\} \in (\mathbb{F}_q^{m \times m})^k$ \cite{SS}.  
	The definitions of SDRM and $\mathbf{D}_{srk}$-code are illustrated in the  following example.
	\begin{example}
		Let $q=2$, $m=2$, and $k=1$. Hence, the message space is $\mathbf{\Pi}=\mathbb{F}_2^{2\times 2}$.
		Consider the following four messages:
		\[
		\mathbf{X}_1=
		\begin{bmatrix}
			0&0\\
			0&0
		\end{bmatrix},
		\quad
		\mathbf{X}_2=
		\begin{bmatrix}
			1&0\\
			0&1
		\end{bmatrix},
		\quad
		\mathbf{X}_3=
		\begin{bmatrix}
			1&0\\
			0&0
		\end{bmatrix},
		\quad
		\mathbf{X}_4=
		\begin{bmatrix}
			0&0\\
			0&1
		\end{bmatrix}.
		\]
		Define a function $f$ by $f(\mathbf{X}_1)=f(\mathbf{X}_2)=0, \qquad f(\mathbf{X}_3)=f(\mathbf{X}_4)=1.$
		 Since $k=1$, the sum-rank distance reduces to the
		rank distance, namely,
		$d_{srk}(\mathbf{X}_i,\mathbf{X}_j)= rk(\mathbf{X}_i-\mathbf{X}_j).$
		The pairwise sum-rank distances are
		\[
		\left(d_{\mathrm{srk}}(\mathbf{X}_i,\mathbf{X}_j)\right)_{i,j\in[4]}
		=
		\begin{bmatrix}
			0&2&1&1\\
			2&0&1&1\\
			1&1&0&2\\
			1&1&2&0
		\end{bmatrix}.
		\]
		From the definition of SDRM, for $t=1$ we obtain
		\[
		\!\mathbf{D}_{srk}^f\!(1,\! \mathbf{X}_1, \dots, \mathbf{X}_4)\!
		=
		\begin{bmatrix}
			0&0&2&2\\
			0&0&2&2\\
			2&2&0&0\\
			2&2&0&0
		\end{bmatrix}.
		\]
		Now, let the redundancy length be $r=1$ and consider
		\[
		\mathbf{C}_1=\mathbf{C}_2=
		\begin{bmatrix}
			0&0\\
			0&0
		\end{bmatrix},
		\qquad
		\mathbf{C}_3=\mathbf{C}_4=
		\begin{bmatrix}
			1&0\\
			0&1
		\end{bmatrix}.
		\]
		From the definition of $\mathbf{D}_{srk}$-code, it can be verified that $\mathcal{C}=\{\mathbf{C}_1,\mathbf{C}_2,\mathbf{C}_3,\mathbf{C}_4\}$ is a $\mathbf{D}_{srk}$-code of length $1$.
	\end{example}  
	The following lemma will be used in the derivation of general Plotkin-like bound in Section \ref{sec_Plotkin_SR}.
	\begin{lem}\cite{SS}
		\label{lemSR}
		Let $\mathcal{P} = \{ \mathbf{P}_1, \mathbf{P}_2, \ldots, \mathbf{P}_M \} \subseteq (\mathbb{F}_q^{m \times m})^k$ be a code of length $r$ and cardinality $M$. Then  
		\begin{equation}\label{eqSR}
		\displaystyle \sum_{ i,j \in [M] : i <j} d_{srk}(\mathbf{P}_i,\mathbf{P}_j) \le \frac{rm}{2q^m}\bigl( M^{2}(q^m-1) - a(q^m-a) \bigr), \text{ where } a = M \mod {q^m}.
		\end{equation}
	\end{lem}
	
	\section{Simplified Plotkin-type Bound for FCCs under b-symbol distance }\label{sec_Plotkin_b}
	A simplified Plotkin-type bound on the optimal redundancy of FCBCs is obtained in this section. Any function $f : \mathbb{F}_q^k \to Im(f)$  partitions the message space $\mathbb{F}_q^k$ into disjoint preimage sets according to the values taken by $f$.  The following theorem gives a Plotkin-type lower bound on the optimal redundancy of FCCs under the b-symbol distance $d_b$ for an arbitrary function $f$. The bound is determined by the size of the preimage sets and the total $b$-symbol distance between pairs of vectors belonging to the same set. The following lemma is used in the proof of the general Plotkin-type bound in Theorem \ref{thm:general_plotkin_b}.
	\begin{lem}\label{lem_bsym}
		For $ \mathbf{u}=(u_0,u_1,...,u_{k-1}), \mathbf{v}=(v_0,v_1,...,v_{k-1}) \in \mathbb{F}_q^k$,  we have 
		\begin{equation}
			\sum_{\mathbf{u}, \mathbf{v} \in \mathbb{F}_q^k} d_{b}(\mathbf{u}, \mathbf{v}) = k\left(q^{2k}-q^{2k-b}\right).
		\end{equation}
	\end{lem}
	\begin{IEEEproof}
		For each $i\in\{0,1,\ldots,k-1\}$, let $B_i(\mathbf{u})=(u_i,u_{i+1},\ldots,u_{i+b-1})$ denote the $i$th cyclic $b$-block of $\mathbf{u}$, where the indices are taken modulo $k$. Then, we have 
		$d_b(\mathbf{u},\mathbf{v}) =\displaystyle \sum_{i=0}^{k-1} \mathbf{1}_{\{B_i(\mathbf{u})\neq B_i(\mathbf{v})\}}$, where $\mathbf{1}_{\{B_i(\mathbf{u})\neq B_i(\mathbf{v})\}}= \begin{cases}
			1, & B_i(\mathbf{u})\neq B_i(\mathbf{v}), \\
			0, & \text{ otherwise }.
		\end{cases}$
		
		Hence,
		\begin{align}\label{eq:sum1}
			\sum_{\mathbf{u},\mathbf{v}\in\mathbb{F}_q^k} d_b(\mathbf{u},\mathbf{v})
			&=
			\sum_{\mathbf{u},\mathbf{v}\in\mathbb{F}_q^k} \sum_{i=0}^{k-1}
			\mathbf{1}_{\{B_i(\mathbf{u})\neq B_i(\mathbf{v})\}} = \sum_{i=0}^{k-1}
			\sum_{\mathbf{u},\mathbf{v}\in\mathbb{F}_q^k}
			\mathbf{1}_{\{B_i(\mathbf{u})\neq B_i(\mathbf{v})\}}.
		\end{align}
		
		For a fixed position $i\in\{0,1,\ldots,k-1\}$, there are $q^b$ possible $b$-blocks, and each $b$-block occurs in exactly $q^{k-b}$ vectors of $\mathbb{F}_q^k$. Therefore, the number of ordered pairs $(\mathbf{u},\mathbf{v})$ satisfying $B_i(\mathbf{u})=B_i(\mathbf{v})$ is $q^b\left(q^{k-b}\right)^2=q^{2k-b}$. Since the total number of ordered pairs is $q^{2k}$, the number of pairs with $B_i(\mathbf{u})\neq B_i(\mathbf{v})$ is $q^{2k}-q^{2k-b}$. That is, $\displaystyle \sum_{\mathbf{u},\mathbf{v}\in\mathbb{F}_q^k}
		\mathbf{1}_{\{B_i(\mathbf{u})\neq B_i(\mathbf{v})\}} = q^{2k}-q^{2k-b}$. Since this count is independent of $i$, from \eqref{eq:sum1} we obtain
		$\displaystyle \sum_{\mathbf{u},\mathbf{v}\in\mathbb{F}_q^k} d_b(\mathbf{u},\mathbf{v})= k\left(q^{2k}-q^{2k-b}\right)$.
	\end{IEEEproof}
	
	\begin{thm}\label{thm:general_plotkin_b}
		Let $f:\mathbb{F}_q^k  \to \operatorname{Im}(f)$ be a function with $Im(f)=\{f_1,f_2,...,f_E\}$. For each $ i \in [E]$, let $\mathcal{U}_{i} \triangleq \{\mathbf{u} \in \mathbb{F}_q^k  : f(\mathbf{u})=f_i \}$ and $n_i = |\mathcal{U}_{i}|$. Then, the optimal redundancy $ r_{b}^f(q, k, t)$ of an FCC for $f$ under the b-symbol distance satisfies 
		\begin{equation}
			r_{b}^f(q, k, t) \ge \frac{(2t-b+2)\left(q^{2k}-\displaystyle \sum_{i=1}^{E} {n_i}^2 \right)+ \displaystyle \sum_{i=1}^{E} \sum_{\mathbf{u},\mathbf{v} \in \mathcal{U}_{i} } d_{b}(\mathbf{u}, \mathbf{v})}{q^{2k}(1-q^{-b})} - k.
		\end{equation}
	\end{thm}
	\begin{IEEEproof}
		Consider the message set $\mathbb{F}_q^k = \{\mathbf{u}_i : i \in [q^k] \}$. Let $\mathbf{B}_f^{(1)}(t, \mathbf{u}_1, \dots, \mathbf{u}_{q^k})$ be the BDRM of a function $f:\mathbb{F}_q^k \to \operatorname{Im}(f)$ given in \eqref{BDRM1}. This BDRM is denoted by $\mathbf{B}$ throughout this proof. Let $\mathcal{P} = \{ \mathbf{p}_1, \mathbf{p}_2, \ldots, \mathbf{p}_{q^k} \} \subseteq \mathbb{F}_q^r$ be a $\mathbf{B}_b$-code corresponding to the BDRM $\mathbf{B}$. From the definition of BDRM, it follows that $$ d_{b}(\mathbf{p}_i, \mathbf{p}_j) \geq [\mathbf{B}]_{ij}, \text{ for all } i, j \in [q^k]. $$  Therefore, we have 
		\begin{equation}\label{eq:sum}
			\sum_{i,j \in [q^k] : i <j} [\mathbf{B}]_{ij} \le \sum_{i,j \in [q^k] : i <j} d_b(\mathbf{p}_i, \mathbf{p}_j).
		\end{equation}
		Combining (\ref{eq:Plotkin}) in Lemma \ref{lem:Plotkin} and (\ref{eq:sum}), we obtain
		\begin{equation}\label{eq:Plotkin1}
			\sum_{i,j \in [q^k] : i <j} [\mathbf{B}]_{ij} \le \frac{rq^{2k}(1-q^{-b})}{2}. 
		\end{equation}
		
		Next, we find a lower bound on $\displaystyle \sum_{i,j \in [q^k] : i <j} [\mathbf{B}]_{ij}$. Let $Im(f)=\{f_1,f_2,...,f_E\}$. We define the sets $\mathcal{U}_{i} \triangleq \{\mathbf{u} \in \mathbb{F}_q^k : f(\mathbf{u})=f_i \}$ and $n_i = |\mathcal{U}_{i}|$, $\forall i \in [E]$. Clearly, $\displaystyle \sum_{i=1}^{E} n_i = q^k$. Therefore, $\displaystyle \sum_{i,j \in [q^k] : i <j} [\mathbf{B}]_{ij}$ can be written as 
		$$ \sum_{i,j \in [q^k] : i <j} [\mathbf{B}]_{ij} = \frac{1}{2} \sum_{i=1}^{E} \sum_{\mathbf{u} \in \mathcal{U}_{i} } \sum_{\mathbf{v} \in \mathbb{F}_q^k } [\mathbf{B}(\mathbf{u}, \mathbf{v})],$$ where $[\mathbf{B}(\mathbf{u}, \mathbf{v})]$ denotes the entry of $\mathbf{B}$ corresponding to row $\mathbf{u}$ and column $\mathbf{v}$. This expression can be further simplified using the definition of BDRM as follows.
		
		\begin{align} \label{eq:ub}
			\sum_{i,j \in [q^k] : i <j} [\mathbf{B}]_{ij} & = \frac{1}{2} \sum_{i=1}^{E} \sum_{\mathbf{u} \in \mathcal{U}_{i} } \sum_{\mathbf{v} \in \mathbb{F}_q^k } [B(\mathbf{u}, \mathbf{v})] \nonumber \\ & = \frac{1}{2} \sum_{i=1}^{E} \sum_{\mathbf{u} \in \mathcal{U}_{i} } \sum_{\mathbf{v} \in \mathbb{F}_q^k \backslash \mathcal{U}_{i} } [2t-b+2-d_{b}(\mathbf{u}, \mathbf{v})]^+ \nonumber \\ & \ge \frac{1}{2} \sum_{i=1}^{E} \sum_{\mathbf{u} \in \mathcal{U}_{i} } \sum_{\mathbf{v} \in \mathbb{F}_q^k \backslash \mathcal{U}_{i} } ( 2t-b+2-d_{b}(\mathbf{u}, \mathbf{v}) ) \nonumber \\ & = \frac{2t-b+2}{2} \sum_{i=1}^{E} \left(\sum_{\mathbf{u} \in \mathcal{U}_{i} } \sum_{\mathbf{v} \in \mathbb{F}_q^k \backslash \mathcal{U}_{i} } 1 \right) - \frac{1}{2} \sum_{i=1}^{E} \sum_{\mathbf{u} \in \mathcal{U}_{i} } \sum_{\mathbf{v} \in \mathbb{F}_q^k \backslash \mathcal{U}_{i} } d_{b}(\mathbf{u}, \mathbf{v}).  
		\end{align}
		
		Since $\displaystyle \sum_{\mathbf{u} \in \mathcal{U}_{i} } \sum_{\mathbf{v} \in \mathbb{F}_q^k \backslash \mathcal{U}_{i} } 1 = n_i (q^k-n_i)$ and $\displaystyle \sum_{i=1}^{E} n_i =q^k$, (\ref{eq:ub}) can be written as 
		\begin{align}\label{eq:inter_class}
			\sum_{i,j \in [q^k] : i <j} [\mathbf{B}]_{ij} & \ge \frac{2t-b+2}{2} \sum_{i=1}^{E} \left(n_i (q^k-n_i) \right) - \frac{1}{2} \sum_{i=1}^{E} \sum_{\mathbf{u} \in \mathcal{U}_{i} } \sum_{\mathbf{v} \in \mathbb{F}_q^k \backslash \mathcal{U}_{i} } d_{b}(\mathbf{u}, \mathbf{v}) \nonumber \\ & = \frac{2t-b+2}{2}  \left(q^{2k}-\sum_{i=1}^{E} {n_i}^2 \right) - \frac{1}{2} \sum_{i=1}^{E} \sum_{\mathbf{u} \in \mathcal{U}_{i} } \sum_{\mathbf{v} \in \mathbb{F}_q^k \backslash \mathcal{U}_{i} } d_{b}(\mathbf{u}, \mathbf{v}) \nonumber \\ & = \frac{2t-b+2}{2}  \left(q^{2k}-\sum_{i=1}^{E} {n_i}^2 \right) - \frac{1}{2}  \left(\sum_{\mathbf{u}, \mathbf{v} \in \mathbb{F}_q^k} d_{b}(\mathbf{u}, \mathbf{v}) - \sum_{i=1}^{E} \sum_{\mathbf{u},\mathbf{v} \in \mathcal{U}_{i} } d_{b}(\mathbf{u}, \mathbf{v}) \right).
		\end{align}
		From Lemma \ref{lem_bsym}, we have  
		\begin{equation}\label{eq:xy}
			\sum_{\mathbf{u}, \mathbf{v} \in \mathbb{F}_q^k} d_{b}(\mathbf{u}, \mathbf{v}) = k\left(q^{2k}-q^{2k-b}\right).
		\end{equation}
		Substituting (\ref{eq:xy}) in (\ref{eq:inter_class}), we obtain
		\begin{equation}\label{eq:Plotkin2}
			\sum_{i,j \in [q^k] : i <j} [\mathbf{B}]_{ij} \ge \frac{2t-b+2}{2}  \left(q^{2k}-\sum_{i=1}^{E} {n_i}^2 \right) - \frac{k\left(q^{2k}-q^{2k-b}\right)}{2}   + \frac{1}{2} \sum_{i=1}^{E} \sum_{\mathbf{u},\mathbf{v} \in \mathcal{U}_{i} } d_{b}(\mathbf{u}, \mathbf{v}). 
		\end{equation}
		Therefore, from (\ref{eq:Plotkin1}) and (\ref{eq:Plotkin2}), we get 
		\begin{equation}\label{eq:Plotkin3}
			\frac{2t-b+2}{2}  \left(q^{2k}-\sum_{i=1}^{E} {n_i}^2 \right) - \frac{k\left(q^{2k}-q^{2k-b}\right)}{2}   + \frac{1}{2} \sum_{i=1}^{E} \sum_{\mathbf{u},\mathbf{v} \in \mathcal{U}_{i} } d_{b}(\mathbf{u}, \mathbf{v}) \le \frac{rq^{2k}(1-q^{-b})}{2}.
		\end{equation}
		Rearranging (\ref{eq:Plotkin3}), we obtain
		\begin{equation}\label{eq:Plotkin4}
			r \ge \frac{(2t-b+2)\left(q^{2k}-\displaystyle \sum_{i=1}^{E} {n_i}^2 \right)+ \displaystyle \sum_{i=1}^{E} \sum_{\mathbf{u},\mathbf{v} \in \mathcal{U}_{i} } d_{b}(\mathbf{u}, \mathbf{v})}{q^{2k}(1-q^{-b})} - k.
		\end{equation}
	\end{IEEEproof}

	The known Plotkin-type bound on the optimal redundancy of FCCs under the $b$-symbol distance, applicable to arbitrary function, is stated in the following lemma.
	\begin{lem}\cite{LBWY}\label{lem:Lenz}
		Let $\mathbf{B}_f^{(1)}(t, \mathbf{u}_1, \dots, \mathbf{u}_{q^k})$ denote the BDRM associated with a function $f$ under the $b$-symbol  distance. Then 
		\begin{equation}
			r_{b}^f(q, k, t) \geq \frac{2q^b}{(q^b - 1)q^{2k}} \sum_{i,j, \, i<j} [\mathbf{B}_f^{(1)}(t, \mathbf{u}_1, \dots, \mathbf{u}_{q^k})]_{ij}.
		\end{equation}
	\end{lem}
	
	From Lemma \ref{lem:Lenz} and the definition of BDRM, it is clear that the existing bound depends on the distance requirements for pairs of message vectors having different function values. Hence, its evaluation can become computationally expensive when the message space is large. In contrast, the bound in Theorem \ref{thm:general_plotkin_b} only requires the sizes of the preimage sets and the sums of pairwise distances between vectors within each preimage set. These quantities are comparatively easier to compute, making the proposed bound more computationally efficient for larger message spaces.
	
	\subsection{Plotkin-type bound for linear functions under $b$-symbol distance}
	We now simplify the general Plotkin-type bound stated in Theorem \ref{thm:general_plotkin_b} for surjective linear functions under the $b$-symbol distance. A function $f : \mathbb{F}_q^{k} \to \mathbb{F}_q^{l} $ is said to be linear if it satisfies the following condition: $$f(\alpha  \mathbf{x} + \beta  \mathbf{y}) = \alpha  f(\mathbf{x}) +\beta f(\mathbf{y}), \forall\hspace{0.1cm} \mathbf{x},\mathbf{y} \in \mathbb{F}_q^{k} \text{ and } \alpha , \beta \in  \mathbb{F}_q.$$ We assume that the linear functions under consideration are surjective. The kernel of $f$, denoted by $ker(f)$, is defined as $ker(f) \triangleq \{\mathbf{x} \in \mathbb{F}_q^{k} : f(\mathbf{x}) =0 \}$. The set of all cosets of $ker(f)$ is denoted by $\mathbb{F}_q^{k}/ker(f)$. Since $f$ is linear, these cosets form a partition of $\mathbb{F}_q^{k}$ where each coset consists precisely of the message vectors that yield the same function value. Therefore, $E=|Im(f)|=q^l$, $\mathcal{U}_{i} = \mathbf{u}_i + ker(f)$ for some $\mathbf{u}_i$ and
	\begin{equation}\label{eq:linear1}
		n_i=|\mathcal{U}_{i}|={|q}^{k-l}, \forall i \in [E].
	\end{equation}  
	Consider $\mathbf{u},\mathbf{v} \in \mathcal{U}_{i}$. We can write $\mathbf{u}=\mathbf{u}_i+\mathbf{a}$ and $\mathbf{v}=\mathbf{u}_i+\mathbf{b}$, where $\mathbf{a}, \mathbf{b} \in ker(f)$. For  $b$-symbol distance we have, $d_{b}(\mathbf{u}, \mathbf{v})=d_{b}(\mathbf{u}_i+\mathbf{a}, \mathbf{u}_i+\mathbf{b})=d_{b}(\mathbf{a}, \mathbf{b})$. Moreover, the mapping $(\mathbf{a}, \mathbf{b}) \rightarrow (\mathbf{u}_i+\mathbf{a},\mathbf{u}_i+\mathbf{b})$ is a bijection from $ker(f) \times ker(f)$ to $\mathcal{U}_{i} \times \mathcal{U}_{i}$. Therefore, we have 
	\begin{equation}\label{eq:linear2}
		\sum_{\mathbf{u},\mathbf{v} \in \mathcal{U}_{i} } d_{b}(\mathbf{u}, \mathbf{v}) = \sum_{\mathbf{a},\mathbf{b} \in ker(f) } d_{b}(\mathbf{a}, \mathbf{b}), \forall i \in [E].
	\end{equation}
	
	 For $b$-symbol distance we have $d_{b}(\mathbf{u}, \mathbf{v})=d_{b}(\mathbf{u}-\mathbf{v},\mathbf{0})=w_b(\mathbf{u}-\mathbf{v})$. For every fixed $\mathbf{u} \in ker(f)$, as $\mathbf{v}$ ranges over $ker(f)$, $\mathbf{w}=\mathbf{u}-\mathbf{v}$ also ranges over $ker(f)$ exactly once. Therefore,
	\begin{equation}\label{eq:linear3}  
		\sum_{\mathbf{u},\mathbf{v} \in ker(f) } d_{b}(\mathbf{u}, \mathbf{v})= \displaystyle \sum_{\mathbf{u},\mathbf{v} \in ker(f) } w_{b}(\mathbf{u}-\mathbf{v}) =|ker(f)| \displaystyle \sum_{\mathbf{w}\in ker(f)}w_{b}(\mathbf{w}) = {q}^{k-l} s, 
	\end{equation}
	where $s=\sum_{\mathbf{w}\in ker(f)}w_{b}(\mathbf{w})$.

	Substituting $E=q^l$, (\ref{eq:linear1}) and (\ref{eq:linear3}) in Theorem \ref{thm:general_plotkin_b}, we obtain a simplified lower bound on the optimal redundancy of FCCs for linear functions as follows.  
	\allowdisplaybreaks
	\begin{align}
		r_{b}^f(q, k, t) & \ge 
		 \frac{(2t-b+2)\left(q^{2k}-q^l q^{2(k-l)} \right)+ q^l q^{k-l}s}{q^{2k}(1-q^{-b})} - k \nonumber 
		& = \frac{(2t-b+2)\left(q^{2k}-q^{2k-l} \right)+ q^{k}s}{q^{2k}(1-q^{-b})} - k \nonumber \\ & = \frac{(2t-b+2)\left(1-q^{-l} \right)}{(1-q^{-b})}+\frac{s}{q^k(1-q^{-b})} - k.
	\end{align} 
	The obtained lower bound on the optimal redundancy of FCCs for linear functions is stated in the following corollary.  
	\begin{cor}\label{cor_linear}
		Let $f : \mathbb{F}_q^{k} \to \mathbb{F}_q^{l} $ be a surjective linear function with $ker(f) \triangleq \{\mathbf{u} \in \mathbb{F}_q^{k} : f(\mathbf{u}) =\mathbf{0} \}$. Then, the optimal redundancy $ r_{b}^f(q, k, t)$ of an FCBC for $f$ satisfies
		\begin{equation}
			r_{b}^f(q, k, t) \ge  \frac{(2t-b+2)\left(1-q^{-l} \right)}{(1-q^{-b})}+\frac{s}{q^k(1-q^{-b})} - k. 
		\end{equation} 
	\end{cor}
	The obtained bound for linear functions in Corollary \ref{cor_linear} is same as the simplified bound for linear functions in \cite{SR}. Thus, the bound in \cite{SR} is subsumed as a special case of the proposed bound in Theorem \ref{thm:general_plotkin_b}.
	
	\section{Simplified Plotkin-type Bound for FCCs under sum-rank distance }\label{sec_Plotkin_SR}
	In this section, we first obtain a simplified Plotkin-type bound on the optimal redundancy of FCSRCs. Any function $f:\mathbb{F}_q^{m\times m})^k\to\operatorname{Im}(f)$ partitions the message space $\mathbb{F}*q^{m\times m})^k$ into disjoint preimage sets according to the function values. The following theorem gives a Plotkin-type lower bound on the optimal redundancy of FCSRCs for an arbitrary function $f$. The bound depends on the sizes of the preimage sets and the total sum-rank distance between pairs of message vectors in the same preimage set. The following lemma is used in the proof of the general Plotkin-type bound in Theorem \ref{thm:general_plotkin_SR}.
	\begin{lem}\label{lem_SR}
		Let $q$ be a prime power and let $\mathbf{\Pi}=(\mathbb F_q^{m\times m})^k$. For $ \mathbf{X}=(\mathbf{X}_1,\mathbf{X}_2,...,\mathbf{X}_k), \mathbf{Y}=(\mathbf{Y}_1,\mathbf{Y}_2,...,\mathbf{Y}_k) \in \mathbf{\Pi}$,  we have 
		\begin{equation}
			\sum_{\mathbf{X}, \mathbf{Y} \in (\mathbb{F}_q^{m \times m})^k} d_{srk}(\mathbf{X}, \mathbf{Y}) = kq^{m^2(2k-1)} \sum_{r=0}^{m}
			r\prod_{j=0}^{r-1} \frac{(q^m-q^j)^2}{q^r-q^j} \le km(q^m-1)q^{2km^2-m}.
		\end{equation}
	\end{lem}
	\begin{IEEEproof} 
		By the definition of the sum-rank distance,
		\begin{equation}\label{eq_a}
			\sum_{\mathbf{X},\mathbf{Y} \in\mathbf{\Pi}}
			d_{\mathrm{srk}}(\mathbf{X},\mathbf{Y})
			=
			\sum_{\mathbf{X},\mathbf{Y} \in\mathbf{\Pi}}
			\sum_{i=1}^{k}\operatorname{rk}(\mathbf{X}_i-\mathbf{Y}_i)\
			=
			\sum_{i=1}^{k}
			\sum_{\mathbf{X},\mathbf{Y} \in\mathbf{\Pi}}
			\operatorname{rk}(\mathbf{X}_i-\mathbf{Y}_i).
		\end{equation}
		Since all $k$ matrix positions contribute the same amount to the total sum, it suffices to consider the contribution from the first coordinate. For a fixed pair $\mathbf{X}_1,\mathbf{Y}_1\in\mathbb F_q^{m\times m}$, there are $q^{m^2(k-1)}$ possible choices for the remaining $k-1$ coordinates of $\mathbf{X}$, and independently $q^{m^2(k-1)}$
		possible choices for the remaining $k-1$ coordinates of $\mathbf{Y}$. Therefore,
		\begin{equation}\label{eq_b}
			\sum_{\mathbf{X},\mathbf{Y} \in \mathbf{\Pi}}
			rk(\mathbf{X}_1-\mathbf{Y}_1) = q^{2m^2(k-1)}
			\sum_{\mathbf{A},\mathbf{B}\in\mathbb F_q^{m\times m}}
			rk(\mathbf{A}-\mathbf{B}).
		\end{equation}
		
		For every fixed $\mathbf{C}\in\mathbb F_q^{m\times m}$, the equation $ \mathbf{A}-\mathbf{B}=\mathbf{C} $
		has exactly \(q^{m^2}\) ordered solutions \((\mathbf{A},\mathbf{B})\), since $\mathbf{B}$ can be chosen arbitrarily and then $\mathbf{A}=\mathbf{C}+\mathbf{B}$ is uniquely determined. Therefore,
		\begin{equation}\label{eq_c}
			\sum_{\mathbf{A},\mathbf{B}\in\mathbb F_q^{m\times m}}
			rk(\mathbf{A}-\mathbf{B}) = q^{m^2} \sum_{\mathbf{C}\in\mathbb F_q^{m\times m}} rk(\mathbf{C}).
		\end{equation}
		
		Combining \eqref{eq_a}, \eqref{eq_b} and \eqref{eq_c}, we obtain
		\begin{equation}\label{eq_d}
			\sum_{\mathbf{X},\mathbf{Y} \in \mathbf{\Pi}} d_{\mathrm{srk}}(\mathbf{X},\mathbf{Y}) = kq^{2m^2(k-1)}q^{m^2} \sum_{\mathbf{C} \in\mathbb F_q^{m\times m}} rk(\mathbf{C}) = kq^{m^2(2k-1)} \sum_{\mathbf{C}\in\mathbb F_q^{m\times m}} rk(\mathbf{C}).
		\end{equation}
		The number of $m\times m$ matrices over $\mathbb F_q$ having rank $r$ is
		given by $N_r=\displaystyle \prod_{j=0}^{r-1} \frac{(q^m-q^j)^2}{q^r-q^j}.$ Therefore,
		\begin{equation}\label{eq_e}
			\sum_{\mathbf{C}\in\mathbb F_q^{m\times m}}rk(\mathbf{C})=
			\sum_{r=0}^{m}rN_r \sum_{r=0}^{m} r\prod_{j=0}^{r-1} \frac{(q^m-q^j)^2}{q^r-q^j}
		\end{equation}
		
		Substituting \eqref{eq_e} in \eqref{eq_d} gives
		\begin{equation}
			\sum_{\mathbf X,\mathbf Y \in\mathbf{\Pi}} d_{\mathrm{srk}}(\mathbf X,\mathbf Y)= kq^{m^2(2k-1)} \sum_{r=0}^{m} r\prod_{j=0}^{r-1}
			\frac{(q^m-q^j)^2}{q^r-q^j}.
		\end{equation}
		
		Under the natural $\mathbb{F}_q$-linear isomorphism $\Phi:(\mathbb{F}_q^{m\times m})^k\to\mathbb{F}_{q^m}^{km}$
		induced by a fixed basis of $\mathbb{F}_{q^m}$, the sum-rank weight satisfies $srk(\mathbf{U})\leq w_H(\Phi(\mathbf{U}))$, where $w_H$ denotes the Hamming weight, for every $\mathbf{U}\in(\mathbb{F}_q^{m\times m})^k$. Since $\Phi$ is
		bijective, this yields $ \displaystyle
		\sum_{\mathbf{U}\in(\mathbb{F}_q^{m\times m})^k}
		srk(\mathbf{u}) \leq \sum_{\mathbf{y}\in\mathbb{F}_{q^m}^{km}}w_H(\mathbf{y})$ \cite{SS}. Since the sum of Hamming weights of all vectors of length $km$ over $\mathbb{F}_{q^m}$, $\sum_{\mathbf{y}\in\mathbb{F}_{q^m}^{km}}w_H(\mathbf{y}) = km q^{m(km-1)}(q^m-1)$, we obtain
		\begin{align}
			\sum_{\mathbf{X},\mathbf{Y}\in (\mathbb{F}_q^{m\times m})^k}
			d_{\mathrm{srk}}(\mathbf{X},\mathbf{Y})&= q^{km^2}
			\sum_{\mathbf{U}\in (\mathbb{F}_q^{m\times m})^k} srk(\mathbf{U})
			\leq q^{km^2} \sum_{\mathbf{y}\in\mathbb{F}_{q^m}^{km}}
			w_H(\mathbf{y}) = q^{km^2}km q^{m(km-1)}(q^m-1)\nonumber \\
			&= km(q^m-1)q^{2km^2-m}.
		\end{align}
	\end{IEEEproof}
	
	\begin{thm}\label{thm:general_plotkin_SR}
		Let $f:(\mathbb{F}_q^{m\times m})^k\longrightarrow \operatorname{Im}(f)$ be a function with $Im(f)=\{f_1,f_2,...,f_E\}$. For each $ i \in [E]$, let $\mathcal{U}_{i} \triangleq \{\mathbf{U} \in (\mathbb{F}_q^{m\times m})^k  : f(\mathbf{U})=f_i \}$ and $n_i = |\mathcal{U}_{i}|$. Then, the optimal redundancy $ r_{srk}^f(m, k, t)$ of an FCSRC for $f$  satisfies 
		\begin{align}
			r_{srk}^f(m, k, t) & \ge \frac{(2t+1)\left(q^{2m^2k}-\displaystyle \sum_{i=1}^{E} {n_i}^2 \right)+ \displaystyle \sum_{i=1}^{E} \sum_{\mathbf{U},\mathbf{V} \in \mathcal{U}_{i} } d_{srk}(\mathbf{U}, \mathbf{V})}{mq^{2m^2k}(1-q^{-m})} -\frac{ k}{mq^{m^2}(1-q^{-m})}\sum_{r=0}^{m}
			r\prod_{j=0}^{r-1} \frac{(q^m-q^j)^2}{q^r-q^j} \label{eq1} \\ & \ge \frac{(2t+1)\left(q^{2m^2k}-\displaystyle \sum_{i=1}^{E} {n_i}^2 \right)+ \displaystyle \sum_{i=1}^{E} \sum_{\mathbf{U},\mathbf{V} \in \mathcal{U}_{i} } d_{srk}(\mathbf{U}, \mathbf{V})}{mq^{2m^2k}(1-q^{-m})} -k \label{eq2}.
		\end{align}
	\end{thm}
	\begin{IEEEproof}
		Consider the message set $(\mathbb{F}_q^{m\times m})^k = \{\mathbf{U}_i : i \in [q^{m^2k}] \}$. Let $\mathbf{D}_{srk}^{f}(t, \mathbf{U}_1, \dots, \mathbf{U}_{q^{m^2k}})$ be the SDRM of a function $f:(\mathbb{F}_q^{m\times m})^k\longrightarrow \operatorname{Im}(f)$ given in \eqref{SDRM}. This SDRM is denoted by $\mathbf{D}$ throughout this proof. Let $\mathcal{P} = \{ \mathbf{P}_1, \mathbf{P}_2, \ldots, \mathbf{P}_{q^{m^2k}} \} \subseteq (\mathbb{F}_q^{m\times m})^r$ be a $\mathbf{D}_{srk}$-code corresponding to the SDRM $\mathbf{D}$. From the definition of SDRM, it follows that $$ d_{srk}(\mathbf{P}_i, \mathbf{P}_j) \geq [\mathbf{D}]_{ij}, \text{ for all } i, j \in [q^{m^2k}]. $$  Therefore, we have 
		\begin{equation}\label{eq:sum_sr}
			\sum_{i,j \in [q^{m^2k}] : i <j} [\mathbf{D}]_{ij} \le \sum_{i,j \in [q^{m^2k}] : i <j} d_{srk}(\mathbf{P}_i, \mathbf{P}_j).
		\end{equation}
		Since $M=q^{m^2k}$, we have $a=0$ in (\ref{eqSR}). Combining (\ref{eqSR}) in Lemma \ref{lemSR} and (\ref{eq:sum_sr}), we obtain
		\begin{equation}\label{eq:Plotkin_SR}
			\sum_{i,j \in [q^{m^2k}] : i <j} [\mathbf{D}]_{ij} \le \frac{rm}{2q^m}\bigl( q^{2m^2k}(q^m-1) \bigr). 
		\end{equation}
		
		We now derive a lower bound on $\displaystyle \sum_{i,j \in [q^{m^2k}] : i <j} [\mathbf{D}]_{ij}$. Let $Im(f)=\{f_1,f_2,...,f_E\}$. We define the sets $\mathcal{U}_{i} \triangleq \{\mathbf{U} \in (\mathbb{F}_q^{m\times m})^k : f(\mathbf{U})=f_i \}$ and $n_i = |\mathcal{U}_{i}|$, $\forall i \in [E]$. Clearly, $\displaystyle \sum_{i=1}^{E} n_i = q^{m^2k}$. Therefore, $\displaystyle \sum_{i,j \in [q^{m^2k}] : i <j} [\mathbf{D}]_{ij}$ can be written as 
		$$ \sum_{i,j \in [q^{m^2k}] : i <j} [\mathbf{D}]_{ij} = \frac{1}{2} \sum_{i=1}^{E} \sum_{\mathbf{U} \in \mathcal{U}_{i} } \sum_{\mathbf{V} \in (\mathbb{F}_q^{m\times m})^k } [\mathbf{D}(\mathbf{U}, \mathbf{V})],$$ where $[\mathbf{D}(\mathbf{U}, \mathbf{V})]$ denotes the entry of $\mathbf{D}$ corresponding to row $\mathbf{U}$ and column $\mathbf{V}$. This expression can be further simplified using the definition of SDRM as follows.
		
		\begin{align} \label{eq:lb_sr}
			\sum_{i,j \in [q^{m^2k}] : i <j} [\mathbf{D}]_{ij} & = \frac{1}{2} \sum_{i=1}^{E} \sum_{\mathbf{U} \in \mathcal{U}_{i} } \sum_{\mathbf{V} \in (\mathbb{F}_q^{m\times m})^k } [D(\mathbf{U}, \mathbf{V})] \nonumber \\ & = \frac{1}{2} \sum_{i=1}^{E} \sum_{\mathbf{U} \in \mathcal{U}_{i} } \sum_{\mathbf{V} \in (\mathbb{F}_q^{m\times m})^k \backslash \mathcal{U}_{i} } [2t+1-d_{srk}(\mathbf{U}, \mathbf{V})]^+ \nonumber \\ & \ge \frac{1}{2} \sum_{i=1}^{E} \sum_{\mathbf{U} \in \mathcal{U}_{i} } \sum_{\mathbf{V} \in (\mathbb{F}_q^{m\times m})^k \backslash \mathcal{U}_{i} } ( 2t+1-d_{srk}(\mathbf{U}, \mathbf{V}) ) \nonumber \\ & = \frac{2t+1}{2} \sum_{i=1}^{E} \left(\sum_{\mathbf{U} \in \mathcal{U}_{i} } \sum_{\mathbf{V} \in (\mathbb{F}_q^{m\times m})^k \backslash \mathcal{U}_{i} } 1 \right) - \frac{1}{2} \sum_{i=1}^{E} \sum_{\mathbf{U} \in \mathcal{U}_{i} } \sum_{\mathbf{V} \in (\mathbb{F}_q^{m\times m})^k \backslash \mathcal{U}_{i} } d_{srk}(\mathbf{U}, \mathbf{V}).  
		\end{align}
		
		Since $\displaystyle \sum_{\mathbf{U} \in \mathcal{U}_{i} } \sum_{\mathbf{V} \in (\mathbb{F}_q^{m\times m})^k \backslash \mathcal{U}_{i} } 1 = n_i (q^{m^2k}-n_i)$ and $\displaystyle \sum_{i=1}^{E} n_i =q^{m^2k}$, (\ref{eq:lb_sr}) can be written as 
		\begin{align}\label{eq:inter_class_sr}
			\sum_{i,j \in [q^{m^2k}] : i <j} [\mathbf{D}]_{ij} & \ge \frac{2t+1}{2} \sum_{i=1}^{E} \left(n_i (q^{m^2k}-n_i) \right) - \frac{1}{2} \sum_{i=1}^{E} \sum_{\mathbf{U} \in \mathcal{U}_{i} } \sum_{\mathbf{V} \in (\mathbb{F}_q^{m\times m})^k \backslash \mathcal{U}_{i} } d_{srk}(\mathbf{U}, \mathbf{V}) \nonumber \\ & = \frac{2t+1}{2}  \left(q^{2m^2k}-\sum_{i=1}^{E} {n_i}^2 \right) - \frac{1}{2} \sum_{i=1}^{E} \sum_{\mathbf{U} \in \mathcal{U}_{i} } \sum_{\mathbf{V} \in (\mathbb{F}_q^{m\times m})^k \backslash \mathcal{U}_{i} } d_{srk}(\mathbf{U}, \mathbf{V}) \nonumber \\ & = \frac{2t+1}{2}  \left(q^{2m^2k}-\sum_{i=1}^{E} {n_i}^2 \right) - \frac{1}{2}  \left(\sum_{\mathbf{U}, \mathbf{V} \in (\mathbb{F}_q^{m\times m})^k} d_{srk}(\mathbf{U}, \mathbf{V}) - \sum_{i=1}^{E} \sum_{\mathbf{U},\mathbf{V} \in \mathcal{U}_{i} } d_{srk}(\mathbf{U}, \mathbf{V}) \right).
		\end{align}
		From Lemma \ref{lem_SR}, we have  
		\begin{equation}\label{eq:uv_SR}
			\sum_{\mathbf{U}, \mathbf{V} \in (\mathbb{F}_q^{m\times m})^k} d_{srk}(\mathbf{U}, \mathbf{V}) = kq^{m^2(2k-1)} \sum_{r=0}^{m}
			r\prod_{j=0}^{r-1} \frac{(q^m-q^j)^2}{q^r-q^j} \le km(q^m-1)q^{2km^2-m}.
		\end{equation}
		Substituting  the equality expression in (\ref{eq:uv_SR}) in (\ref{eq:inter_class_sr}), we obtain
		\begin{equation}\label{eq:PlotkinSR}
			\sum_{i,j \in [q^{m^2k}] : i <j} [\mathbf{D}]_{ij} \ge \frac{2t+1}{2}  \left(q^{2m^2k}-\sum_{i=1}^{E} {n_i}^2 \right) - \frac{kq^{m^2(2k-1)} }{2} \sum_{r=0}^{m}
			r\prod_{j=0}^{r-1} \frac{(q^m-q^j)^2}{q^r-q^j}   + \frac{1}{2} \sum_{i=1}^{E} \sum_{\mathbf{U},\mathbf{V} \in \mathcal{U}_{i} } d_{srk}(\mathbf{U}, \mathbf{V}). 
		\end{equation}
		Therefore, from (\ref{eq:Plotkin_SR}) and (\ref{eq:PlotkinSR}), we get 
		\begin{align}\label{eq:Plotkin3_SR}
			\frac{2t+1}{2}  \left(q^{2m^2k}-\sum_{i=1}^{E} {n_i}^2 \right) - \frac{kq^{m^2(2k-1)} }{2} \sum_{r=0}^{m}
			r\prod_{j=0}^{r-1} \frac{(q^m-q^j)^2}{q^r-q^j}   + \frac{1}{2} \sum_{i=1}^{E} \sum_{\mathbf{U},\mathbf{V} \in \mathcal{U}_{i} } d_{srk}(\mathbf{U}, \mathbf{V}) \nonumber \\ \le  \frac{rm}{2q^m}\bigl( q^{2m^2k}(q^m-1).
		\end{align}
		Rearranging (\ref{eq:Plotkin3_SR}), we obtain
		\begin{equation}\label{eq:Plotkin4_SR}
			r \ge \frac{(2t+1)\left(q^{2m^2k}-\displaystyle \sum_{i=1}^{E} {n_i}^2 \right)+ \displaystyle \sum_{i=1}^{E} \sum_{\mathbf{U},\mathbf{V} \in \mathcal{U}_{i} } d_{srk}(\mathbf{U}, \mathbf{V})}{mq^{2m^2k}(1-q^{-m})} -\frac{ k}{mq^{m^2}(1-q^{-m})}\sum_{r=0}^{m}
			r\prod_{j=0}^{r-1} \frac{(q^m-q^j)^2}{q^r-q^j}.
		\end{equation}
		Substituting  the inequality expression in (\ref{eq:uv_SR}) in (\ref{eq:inter_class_sr}), we obtain
		\begin{equation}\label{eq:PlotkinSR2}
			\sum_{i,j \in [q^{m^2k}] : i <j} [\mathbf{D}]_{ij} \ge \frac{2t+1}{2}  \left(q^{2m^2k}-\sum_{i=1}^{E} {n_i}^2 \right) - km(q^m-1)q^{2km^2-m}   + \frac{1}{2} \sum_{i=1}^{E} \sum_{\mathbf{U},\mathbf{V} \in \mathcal{U}_{i} } d_{srk}(\mathbf{U}, \mathbf{V}). 
		\end{equation}
		Therefore, from (\ref{eq:Plotkin_SR}) and (\ref{eq:PlotkinSR2}), we get 
		\begin{align}\label{eq:Plotkin3_SR2}
			\frac{2t+1}{2}  \left(q^{2m^2k}-\sum_{i=1}^{E} {n_i}^2 \right) - km(q^m-1)q^{2km^2-m}   + \frac{1}{2} \sum_{i=1}^{E} \sum_{\mathbf{U},\mathbf{V} \in \mathcal{U}_{i} } d_{srk}(\mathbf{U}, \mathbf{V}) \le  \frac{rm}{2q^m}\bigl( q^{2m^2k}(q^m-1).
		\end{align}
		Rearranging (\ref{eq:Plotkin3_SR2}), we obtain
		\begin{equation}\label{eq:Plotkin4_SR2}
			r \ge \frac{(2t+1)\left(q^{2m^2k}-\displaystyle \sum_{i=1}^{E} {n_i}^2 \right)+ \displaystyle \sum_{i=1}^{E} \sum_{\mathbf{U},\mathbf{V} \in \mathcal{U}_{i} } d_{srk}(\mathbf{U}, \mathbf{V})}{mq^{2m^2k}(1-q^{-m})} -k.
		\end{equation}
	\end{IEEEproof}
	
	The known Plotkin-type bound on the optimal redundancy of FCCs under the sum-rank distance, applicable to arbitrary function, is stated in the following lemma.
	\begin{lem}\cite{SS}\label{lem:SS}
		Let $\mathbf{D}_{srk}^{f}(t, \mathbf{U}_1, \dots, \mathbf{U}_{q^{m^2k}})$ denote the SDRM associated with a function $f$ under the sum-rank distance. Then 
		\begin{equation}
			r_{srk}^f(m, k, t) \geq \frac{2q^m}{m(q^{2m^2k}(q^m-1))}\sum_{1 \le i < j \le q^{m^2k}} [\mathbf{D}_{srk}^{f}(t, \mathbf{U}_1, \dots, \mathbf{U}_{q^{m^2k}})]_{ij}.
		\end{equation}
	\end{lem}
	
	From Lemma \ref{lem:SS} and the definition of SDRM, the existing bound requires the distance requirements between pairs of message vectors corresponding to different function values. As the size of the message space increases, determining these distance requirements can become computationally expensive. On the other hand, the bound in Theorem \ref{thm:general_plotkin_SR} avoids this requirement and depends only on the sizes of the preimage sets and the sums of pairwise distances between vectors within each preimage set. Since these quantities are comparatively easier to compute, the proposed bound is more computationally efficient for larger message spaces.
	
	\subsection{Plotkin-type bound for linear functions under sum-rank distance}
	We now simplify the general Plotkin-type bound stated in Theorem \ref{thm:general_plotkin_SR} for linear functions under the sum-rank distance. A function $f:(\mathbb{F}_q^{m\times m})^k\to(\mathbb{F}_q^{m \times m})^l $ is said to be linear if it satisfies the following condition: $$f(\alpha  \mathbf{X} + \beta  \mathbf{Y}) = \alpha  f(\mathbf{X}) +\beta f(\mathbf{Y}), \forall\hspace{0.1cm} \mathbf{X},\mathbf{Y} \in (\mathbb{F}_q^{m\times m})^k \text{ and } \alpha , \beta \in  \mathbb{F}_q.$$ We assume that the linear functions under consideration are surjective. The kernel of $f$, denoted by $ker(f)$, is defined as $ker(f) \triangleq \{\mathbf{X} \in (\mathbb{F}_q^{m\times m})^k : f(\mathbf{X}) =0 \}$. The set of all cosets of $ker(f)$ is denoted by $(\mathbb{F}_q^{m\times m})^k/ker(f)$. Since $f$ is linear, these cosets form a partition of $(\mathbb{F}_q^{m\times m})^k$ where each coset consists precisely of the message vectors that yield the same function value. Therefore, $E=|Im(f)|=q^{m^2l}$, $\mathcal{U}_{i} = \mathbf{U}_i + ker(f)$ for some $\mathbf{U}_i$ and
	\begin{equation}\label{eq:linear1_SR}
		n_i=|\mathcal{U}_{i}|={q}^{m^2(k-l)}, \forall i \in [E].
	\end{equation}  
	Consider $\mathbf{U},\mathbf{V} \in \mathcal{U}_{i}$. We can write $\mathbf{U}=\mathbf{U}_i+\mathbf{A}$ and $\mathbf{V}=\mathbf{U}_i+\mathbf{B}$, where $\mathbf{A}, \mathbf{B} \in ker(f)$. For sum-rank distance we have, $d_{srk}(\mathbf{U}, \mathbf{V})=d_{srk}(\mathbf{U}_i+\mathbf{A}, \mathbf{U}_i+\mathbf{B})=d_{srk}(\mathbf{A}, \mathbf{B})$. Moreover, the mapping $(\mathbf{A}, \mathbf{B}) \rightarrow (\mathbf{U}_i+\mathbf{A},\mathbf{U}_i+\mathbf{B})$ is a bijection from $ker(f) \times ker(f)$ to $\mathcal{U}_{i} \times \mathcal{U}_{i}$. Therefore, we have 
	\begin{equation}\label{eq:linear2_SR}
		\sum_{\mathbf{U},\mathbf{V} \in \mathcal{U}_{i} } d_{srk}(\mathbf{U}, \mathbf{V}) = \sum_{\mathbf{A},\mathbf{B} \in ker(f) } d_{srk}(\mathbf{A}, \mathbf{B}), \forall i \in [E].
	\end{equation}
	
	For sum-rank distance we have $d_{srk}(\mathbf{U}, \mathbf{V})=d_{srk}(\mathbf{U}-\mathbf{V},\mathbf{0})=w_{srk}(\mathbf{U}-\mathbf{V})$. For every fixed $\mathbf{U} \in ker(f)$, as $\mathbf{V}$ ranges over $ker(f)$, $\mathbf{W}=\mathbf{U}-\mathbf{V}$ also ranges over $ker(f)$ exactly once. Therefore,
	\begin{equation}\label{eq:linear3_SR}  
		\sum_{\mathbf{U},\mathbf{V} \in ker(f) } d_{srk}(\mathbf{U}, \mathbf{V})= \displaystyle \sum_{\mathbf{U},\mathbf{V} \in ker(f) } w_{srk}(\mathbf{U}-\mathbf{V}) =|ker(f)| \displaystyle \sum_{\mathbf{W}\in ker(f)}w_{srk}(\mathbf{W}) = {q}^{m^2(k-l)} s, 
	\end{equation}
	where $s=\displaystyle \sum_{\mathbf{W}\in ker(f)}w_{srk}(\mathbf{W})$.

	Substituting $E=q^{m^2l}$, (\ref{eq:linear1_SR}) and (\ref{eq:linear3_SR}) in \eqref{eq1} in Theorem \ref{thm:general_plotkin_SR}, we obtain a simplified lower bound on the optimal redundancy of FCSRCs for linear functions as follows.  
	\allowdisplaybreaks
	\begin{align}
		r_{srk}^f(m, k, t) & \ge \frac{(2t+1)\left(q^{2m^2k}-q^{2m^2k-m^2l}\right)+ q^{m^2l}q^{m^2(k-l)}s}{mq^{2m^2k}(1-q^{-m})} -\frac{ k}{mq^{m^2}(1-q^{-m})}\sum_{r=0}^{m}
		r\prod_{j=0}^{r-1} \frac{(q^m-q^j)^2}{q^r-q^j} \nonumber\\  & = \frac{(2t+1)\left(1-q^{-m^2l} \right)}{m(1-q^{-m})}+\frac{s}{mq^{m^2k}(1-q^{-m})} - \frac{ k}{mq^{m^2}(1-q^{-m})}\sum_{r=0}^{m}
		r\prod_{j=0}^{r-1} \frac{(q^m-q^j)^2}{q^r-q^j}.
	\end{align} 
	Similarly, substituting $E=q^{m^2l}$, (\ref{eq:linear1_SR}) and (\ref{eq:linear3_SR}) in \eqref{eq2} in Theorem \ref{thm:general_plotkin_SR}, we obtain a simplified lower bound as follows.  
	\allowdisplaybreaks
	\begin{equation}
		r_{srk}^f(m, k, t) \ge \frac{(2t+1)\left(1-q^{-m^2l} \right)}{m(1-q^{-m})}+\frac{s}{mq^{m^2k}(1-q^{-m})} - k.
	\end{equation}
	The obtained lower bounds on the optimal redundancy of FCSRCs for linear functions is stated in the following corollary.  
	\begin{cor}\label{cor_linear_SR}
		Let $f : (\mathbb{F}_q^{m\times m})^k\to(\mathbb{F}_q^{m \times m})^l $ be a surjective linear function with $ker(f) \triangleq \{\mathbf{U} \in (\mathbb{F}_q^{m\times m})^k : f(\mathbf{U}) =\mathbf{0} \}$. Then, the optimal redundancy $ r_{srk}^f(m, k, t)$ of an FCSRC for $f$ satisfies
		\begin{align}
			r_{srk}^f(m, k, t) & \ge  \frac{(2t+1)\left(1-q^{-m^2l} \right)}{m(1-q^{-m})}+\frac{s}{mq^{m^2k}(1-q^{-m})} - \frac{ k}{mq^{m^2}(1-q^{-m})}\sum_{r=0}^{m}
			r\prod_{j=0}^{r-1} \frac{(q^m-q^j)^2}{q^r-q^j} \label{eq_linear1} \\ & \ge \frac{(2t+1)\left(1-q^{-m^2l} \right)}{m(1-q^{-m})}+\frac{s}{mq^{m^2k}(1-q^{-m})} - k \label{eq_linear2}.
		\end{align}
	\end{cor}
	The obtained bound for linear functions in \eqref{eq_linear2} in Corollary \ref{cor_linear_SR} is the same as the simplified bound for linear functions in \cite{SS}. Thus, the bound in \cite{SS} is subsumed as a special case of the proposed bound in Theorem \ref{thm:general_plotkin_SR}. Whereas the bound in \eqref{eq_linear1} in Corollary \ref{cor_linear_SR} is a tighter bound for linear functions compared to the bound in \cite{SS}.

	\section{Conclusion}\label{sec_concl}
In this paper, we derived simplified Plotkin-type bounds on the optimal redundancy of function-correcting codes under the $b$-symbol and sum-rank distances. Unlike existing general Plotkin-type bounds, which require the computation of pairwise distances between message vectors corresponding to different function values, the proposed bounds depend only on the sizes of the preimage sets and the sums of pairwise distances among vectors within each preimage set. Consequently, the proposed bounds are considerably easier to evaluate, particularly for large message spaces. The bounds hold for arbitrary functions and, as special cases, subsume the known simplified bounds for linear functions under both the $b$-symbol and sum-rank distances. Furthermore, for linear functions under the sum-rank distance, we proposed a tighter bound than the existing one.

\end{document}